\documentclass[final,3p]{elsarticle}

\graphicspath{{./}{./figures/}{./article_figures/}}

\usepackage{amsmath}
\usepackage{amssymb}
\usepackage{gensymb}
\usepackage[dvipsnames,svgnames,x11names]{xcolor}

\usepackage{float}
\usepackage{tabularx}
\usepackage{subcaption}
\usepackage{hyperref}
\usepackage[switch, modulo, displaymath, mathlines]{lineno}
\newcommand*\patchAmsMathEnvironmentForLineno[1]{
  \expandafter\let\csname old#1\expandafter\endcsname\csname #1\endcsname
  \expandafter\let\csname oldend#1\expandafter\endcsname\csname end#1\endcsname
  \renewenvironment{#1}
  {\linenomath\csname old#1\endcsname}
  {\csname oldend#1\endcsname\endlinenomath}}
  \newcommand*\patchBothAmsMathEnvironmentsForLineno[1]{
  \patchAmsMathEnvironmentForLineno{#1}
  \patchAmsMathEnvironmentForLineno{#1*}}
  \AtBeginDocument{
  \patchBothAmsMathEnvironmentsForLineno{equation}
  \patchBothAmsMathEnvironmentsForLineno{align}
  \patchBothAmsMathEnvironmentsForLineno{flalign}
  \patchBothAmsMathEnvironmentsForLineno{alignat}
  \patchBothAmsMathEnvironmentsForLineno{gather}
  \patchBothAmsMathEnvironmentsForLineno{multline}
}
\hypersetup{
	colorlinks=true,
	linkcolor=blue,
	filecolor=magenta,      
	urlcolor=cyan,
}

\usepackage{placeins} 
\usepackage{multirow}

\usepackage{xifthen}
\usepackage[final]{changes}
\NewCommandCopy{\addedRI}{\added}
\NewCommandCopy{\replacedRI}{\replaced}
\NewCommandCopy{\deletedRI}{\deleted}
\makeatletter
\renewcommand{\addedRI}[1]{%
	\@namedef{Changes@AuthorColor}{red}%
	\colorlet{Changes@Color}{red}%
	\added{#1}%
}
\renewcommand{\replacedRI}[1]{%
	\@namedef{Changes@AuthorColor}{red}%
	\colorlet{Changes@Color}{red}%
	\replaced{#1}%
}
\renewcommand{\deletedRI}[1]{%
	\@namedef{Changes@AuthorColor}{red}%
	\colorlet{Changes@Color}{red}%
	\deleted{#1}%
}
\makeatother
\NewCommandCopy{\addedRII}{\added}
\NewCommandCopy{\replacedRII}{\replaced}
\NewCommandCopy{\deletedRII}{\deleted}
\makeatletter
\renewcommand{\addedRII}[1]{%
	\@namedef{Changes@AuthorColor}{blue}%
	\colorlet{Changes@Color}{blue}%
	\added{#1}%
}
\renewcommand{\replacedRII}[1]{%
	\@namedef{Changes@AuthorColor}{blue}%
	\colorlet{Changes@Color}{blue}%
	\replaced{#1}%
}
\renewcommand{\deletedRII}[1]{%
	\@namedef{Changes@AuthorColor}{blue}%
	\colorlet{Changes@Color}{blue}%
	\deleted{#1}%
}
\makeatother

\NewCommandCopy{\addedRIII}{\added}
\NewCommandCopy{\replacedRIII}{\replaced}
\NewCommandCopy{\deletedRIII}{\deleted}
\makeatletter
\renewcommand{\addedRIII}[1]{%
	\@namedef{Changes@AuthorColor}{green}%
	\colorlet{Changes@Color}{green}%
	\added{#1}%
}
\renewcommand{\replacedRIII}[1]{%
	\@namedef{Changes@AuthorColor}{green}%
	\colorlet{Changes@Color}{green}%
	\replaced{#1}%
}
\renewcommand{\deletedRIII}[1]{%
	\@namedef{Changes@AuthorColor}{green}%
	\colorlet{Changes@Color}{green}%
	\deleted{#1}%
}
\makeatother

\NewCommandCopy{\addedRIV}{\added}
\NewCommandCopy{\replacedRIV}{\replaced}
\NewCommandCopy{\deletedRIV}{\deleted}
\makeatletter
\renewcommand{\addedRIV}[1]{%
	\@namedef{Changes@AuthorColor}{orange}%
	\colorlet{Changes@Color}{orange}%
	\added{#1}%
}
\renewcommand{\replacedRIV}[1]{%
	\@namedef{Changes@AuthorColor}{orange}%
	\colorlet{Changes@Color}{orange}%
	\replaced{#1}%
}
\renewcommand{\deletedRIV}[1]{%
	\@namedef{Changes@AuthorColor}{orange}%
	\colorlet{Changes@Color}{orange}%
	\deleted{#1}%
}
\makeatother

\journal{Elsevier}

\begin{document}

\begin{frontmatter}



\title{Approximation of sea surface velocity field by fitting surrogate two-dimensional flow to scattered measurements }


\author[riteh]{Karlo Jakac}
\ead{karlo.jakac@uniri.hr}
\author[riteh]{Luka Lan\v{c}a}
\ead{luka.lanca@uniri.hr}
\author[cnrm]{Ante Sikirica}
\ead{ante.sikirica@uniri.hr}
\author[riteh]{Stefan Ivi\'c\corref{cor}}
\ead{stefan.ivic@uniri.hr}

\affiliation[riteh]{organization={Faculty of Engineering, University of Rijeka},
	addressline={Vukovarska 58}, 
	city={Rijeka},
	postcode={51000}, 
	country={Croatia}}

\affiliation[cnrm]{organization={Center for Advanced Computing and Modelling, University of Rijeka},
	addressline={Radmile Matejčić 2}, 
	city={Rijeka},
	postcode={51000}, 
	country={Croatia}}

\cortext[cor]{Corresponding author}

\begin{abstract}
In this paper, a rapid approximation method is introduced to estimate the sea surface velocity field based on scattered measurements. The method uses a simplified two-dimensional flow model as a surrogate model, which mimics the real submesoscale flow. The proposed approach treats the interpolation of the flow velocities as an optimization problem, aiming to fit the flow model to the scattered measurements. To ensure consistency between the simulated velocity field and the measured values, the
boundary conditions in the numerical simulations are adjusted during the optimization process. Additionally, the relevance of quantity and quality of the scattered measurements is assessed, emphasizing the importance of the measurement locations within the domain as well as explaining how these measurements contribute to the accuracy and reliability of the sea surface velocity field approximation. The proposed methodology has been successfully tested in both synthetic and real-world scenarios, leveraging measurements obtained from Global Positioning System (GPS) drifters and high-frequency (HF) radar systems. The adaptability of this approach for different domains, measurement types and conditions implies that it is suitable for real-world submesoscale scenarios where only an approximation of the sea surface velocity field is sufficient.
\end{abstract}




\end{frontmatter}


\section{Introduction}

The significance of oceanic data became more evident as scientists established a connection between changes in oceanic circulation patterns and shifts in the global climate. To gather such data, satellite tracking of deployed drifters has emerged as a preferred method due to its cost-effectiveness in providing velocity data, compared to the amount of data collected from ships and ports. Consequently, over the past two decades, the deployment of drifters at various oceanic locations has increased substantially \cite{sombardier1994global}. Between 1994 and 1995, the Gulf of Mexico witnessed the deployment of 300 satellite-tracked drifters. The data collected from this deployment has been thoroughly analyzed and documented in numerous publications \cite{ohlmann2005circulation,dimarco2005statistical}. 

Similarly, satellite-tracked drifters were deployed in the semi-enclosed Adriatic Sea to determine the circulation properties \cite{lacorata2001drifter,falco2000transport}. Over 200 drifters were deployed in the Adriatic Sea between 1990 and 1999, serving both academic and military purposes. The collected data was utilized to reconstruct an Eulerian velocity field of the Adriatic Sea. This involved interpolating velocities at the drifter locations through the use of cubic splines  \cite{poulain2001adriatic}.  Due to the scope and measurement time frame, this type of data can be used to reconstruct surface circulation in relation to wind forcing, river runoff, and bottom topography \cite{ursella2006surface}.

Typically, drifters can be advected across expansive areas exceeding 2000 square kilometers while the precision of the reconstructed velocity field is heavily reliant on the spatial coverage of the drifter paths \cite{toner2001reconstructing}. Reconstruction of surface velocity fields in submesoscale processes, which occur at domain scales ranging from 0.1 to 10 kilometers and time scales ranging from 1 to 100 hours, cannot be resolved with satellite altimeters and observations by ships. A novelty in this area was The Lagrangian Submesoscale Experiment (LASER) which provided measurements of the surface velocity field across the northern Gulf of Mexico with high spatial and temporal resolution \cite{haza2018drogue}.
A central aspect of this experiment involved releasing over 1000 surface drifters in submesoscale domains. Released drifters were predominantly biodegradable and were outfitted with drogues to negate the impact of wind and waves. The deployment density of these drifters facilitates the capture of scales ranging from tens of kilometers down to tens of meters.

In the last decade, interest in Lagrangian data for operational modeling such as search and rescue, the spread of pollutants, and forecasting has increased significantly \cite{chaturvedi2020mathematical,gulakaram2018role}. This type of data makes it possible to define models that can predict the future behavior of surface currents, which in some cases can save lives. Two relevant methods have been explored for the Lagrangian data reconstruction and integration. The initial method relies on estimating velocities along trajectories by calculating the ratio of observed positional changes to time increments \cite{hernandez1995mapping}, and then directly utilizing these velocities to rectify the model outcomes. In contrast, the second method introduces an observational operator founded on the particle advection equation. It adjusts the Eulerian velocity field by minimizing the disparity between observed trajectories and corresponding model predictions \cite{molcard2003assimilation}.

The primary issue with Lagrangian drifter data stems from the fact that the drifters move with the oceanic flow, hence they are not evenly spread out and have a tendency to either cluster or move out of areas of interest. Furthermore, Lagrangian data describes trajectories that vary in space and time whereas e.g. temperature and velocity are of interest. Transitioning from trajectory analysis, the focus in this domain often shifts towards the detailed reconstruction of Eulerian velocity fields, where research papers emphasize the use of drifter data for velocity field reconstruction \cite{rao1981method,eremeev1992reconstruction,cho1998objectively}.

A study by \cite{gonccalves2019reconstruction} estimated the velocity field based on drifter data employing a method that statistically interpolates the collected data in both spatial and temporal domains. The authors concluded that due to the varying distances between the drifters as they move with the flow, the resolution they offer can fluctuate both spatially and temporally, potentially resulting in interpolation errors. Velocity field estimates at similar submesoscales with high resolution are presented in \cite{d2011enhanced,haza2018drogue,novelli2017biodegradable}. \replacedRIV{Another challenge in flow reconstruction is that restricted and potentially corrupted measurements can significantly affect the results, leading to issues such as overfitting and increased noise. Therefore a}{Common methods for flow reconstruction frequently involve employing least squares regression for reconstructing the flow field or employing basic interpolation to identify the minimum-energy solution that aligns with the collected data. Nevertheless, this methodology is susceptible to over-fitting and is prone to noise due to the presence of restricted and potentially corrupted measurements. A} study by \cite{callaham2019robust} introduced a technique for reconstructing flow fields using sparse representation within a library of examples, specifically tailored for structured data with restricted, flawed measurements. This methodology has the potential to be applied to complex flow fields through spatial domain decomposition.

Lagrangian satellite-tracked drifters are driven by ocean currents, which means there is minimal control over the location of the measurements. Furthermore, they frequently offer restricted coverage of a targeted area, and the acquired data may exhibit unpredictable temporal patterns \cite{shu2023characterising}. In contrast, Eulerian velocity data offering global coverage is accessible, yet its utility is restricted by the measurement resolution of satellite altimetry. In addition, altimetry measurements depend on the process of interpolation and smoothing from raw satellite data. The challenge associated with satellite measurements lies in their limited accessibility and occasional atmospheric interference. This has led to the widespread adoption of an alternative method for acquiring surface velocity using high-frequency (HF) radar. This method utilizes electromagnetic waves to measure surface currents in near-real-time and is commonly used for ocean currents validation \cite{solano2018development}. These measurements can be assimilated into a numerical model, as demonstrated by \cite{marmain2014assimilation} in the case of the northwestern Mediterranean Sea. The assimilation process involved adjusting the model's initial conditions and boundary conditions to match the observed surface currents. The study found that assimilating HF-radar data into the model improved the accuracy, particularly in areas where surface currents were difficult to measure. 

While HF-radar technology provides notable advantages, it comes with certain limitations, such as restricted depth penetration, susceptibility to interference, and difficulties in achieving fine spatial resolution. To address these challenges, \cite{inazu2010optimization} used a genetic algorithm to optimize the boundary conditions and physical parameters of the model. The performance of the model, i.e. the optimized depth boundary condition, was compared to Multibeam Bathymetric data. In addition, the sensitivity of the model to various parameters such as bottom friction, eddy viscosity, and the shape of the coastline was discussed. It was determined that bottom friction is the most influential parameter and was hence suggested that it should be carefully calibrated to enhance the precision of the model. Moreover, for the purpose of swift deployment and emergency observations, \cite{yang2024research} utilized a portable underwater profiler (PUP) which integrates characteristics of both profiling buoys and underwater gliders. This enhancement led to improved accuracy in predicting ocean current velocity through the utilization of full-depth-averaged measurements.

Artificial Intelligence (AI) and machine-learning techniques have recently appeared as alternatives to traditional interpolation and optimization methods for flow field predictions, as discussed in \cite{ghalambaz2024forty}. Several studies, such as those by  \cite{aksamit2020machine,grossi2020predicting}, have explored the application of recurrent neural networks to analyze temporal patterns in drifter motion, providing reduced errors in drifter models. Moreover, the authors in \cite{sun2020surrogate,tang2021deep} employed deep learning (DL) for surrogate modeling of fluid flows, offering a fast and straightforward estimation without relying on extensive CFD simulations. Research presented by \cite{hu2024physics} introduced a Physics-Informed Neural Network (PINN) method in conjunction with the characteristic-based split (CBS) method, effectively solving shallow-water equations and rendering it suitable for estimating surface flow fields. In addition, the paper by \cite{zhao2023sea} introduces a method for sea surface reconstruction utilizing marine radar images in conjunction with deep convolutional neural networks (CNN), resulting in significantly higher accuracy than that achieved by conventional spectral analysis methods.

Furthermore, in addressing the challenge of incomplete data in flow measurements, authors in \cite{kazemi2023adaptive} introduced a model based on an adaptive neuro-fuzzy inference system, utilized for estimating missing velocity vectors in fluid dynamics. These AI and machine learning approaches have garnered significant attention in CFD, demonstrating notable success in turbulence modeling, shape optimization, and flow field prediction.

In this study, a novel simulation-based optimization approach is introduced. By combining a simplified two-dimensional surrogate flow model and an optimization algorithm, the boundary conditions of a flow simulation can be adjusted, i.e. the velocity field can be aligned with the velocity values derived from the scattered drifters. The proposed approach leverages computational power to directly derive complete velocity fields from scattered measurements, which often vary in number, thereby eliminating the need for interpolation schemes and the estimation of velocity components through central-valued finite differences of interpolated positions, as seen in previous works. \cite{hansen1996quality, laurindo2017improved}. While this approach enables obtaining the entire velocity field regardless the available number of drifters, the accuracy of the velocity approximation increases with the number of drifters. The importance of such an algorithmic approach is not limited to academic research, as it can be adapted for use in search and rescue scenarios. By accurately estimating the velocity field in a given area, it enables the prediction of the movement of individuals or objects lost at sea. This capability is crucial for planning and executing rescue operations with greater precision and efficiency. Beyond search and rescue operations, the proposed simulation-based fitting approach emerges as a promising alternative to traditional methods of velocity field reconstruction, and has the potential to be applied in diverse applications within the fields of fluid mechanics and ocean engineering.

\section{Surrogate two-dimensional flow model}

Models involving unsteady flows require substantial computational resources, resulting in prolonged simulation times. As this data is rarely of interest in large-scale models, a surrogate flow model based on stationary incompressible flow has been adopted. Our methodology intentionally excludes influential phenomena such as wind, waves, changing tides, and temperature variations present in the sea flow. This deliberate omission positions our model as a tool for rapid approximation of sea surface flow. Despite these simplifications, our simulation-based optimization approach offers several advantages over traditional methods. No extensive data, additional interpolation, or finite difference computation is required, thus considerably reducing the cost and time. Additionally, the approach relies on CFD simulations, enabling the capture of fluid flow physics and providing more accurate velocity fields. The suggested method is particularly valuable for applications requiring a comprehensive understanding of flow characteristics, where changes in flow occur slowly.

\subsection{Stationary 2D flow model}
\label{subsec:Stationary_2D_flow_model}

The proposed stationary flow model describes the motion of fluids at low and medium velocities in a connected computational domain $\Omega \subset \mathbb{R}^2$ based on the incompressible Navier-Stokes equations \cite{gunzburger2012finite, lions1996mathematical, kronbichler2009numerical}:

\begin{equation}
	\rho\left(\mathbf{u}\cdot\nabla\right)\mathbf{u}-\nabla \cdot \left(2\mu\epsilon(\mathbf{u})\right)+\nabla p = \rho\mathbf{f}
	\label{eq:ns_equation}
\end{equation}
\begin{equation}
	\nabla \cdot \mathbf{u} = 0
	\label{eq:compressibility_equation}
\end{equation}

The vector $\mathbf{u}$ represents the fluid velocity, and $p$ denotes the fluid's dynamic pressure. The rank-2 tensor $\epsilon(\mathbf{u}) = \frac{1}{2} \left( \nabla \mathbf{u} + (\nabla \mathbf{u})^\top \right)$ defines the viscous stress tensor for an incompressible Newtonian fluid. Here, $\rho$ denotes fluid density, and $\mu$ represents the fluid's dynamic viscosity. Density remains constant due to the assumption of incompressibility. The term $\mathbf{f}$ specifies external forces acting on the fluid.

To account for the interactions between the observed section of the sea and the broader sea domain, a specific combination of boundary conditions is employed. This combination ensures an adequate representation of the dynamic interactions at the boundary and increases the reliability of the computational model. The proposed model reconstructs the velocity field by generating tangential velocities and pressure values at the boundary. However, these boundary values should not be randomly generated i.e. their range should reflect realistic circumstances in order to recreate the physics of a realistic flow. \addedRIV{According to \protect\cite{cosoli2013surface,notarstefano2008estimation,bolanos2014modelling} , surface velocities in the Adriatic Sea, estimated using radar data, numerical models, and infrared satellite images, showed significant variability, with speeds ranging from below 0.1 m/s to above 0.5 m/s. The authors also noted that the average surface velocities were around 0.1-0.2 m/s in most areas. Due to this high variability in surface currents, the boundary values for pressure and tangential velocity are set to achieve a velocity magnitude of 0.5 m/s within the domain. Since the tangential velocity is prescribed, based on the internal velocity magnitude limit, normal velocity component can be derived and used to calculate the total pressure at the boundary:}
\addedRIV{
\begin{equation}
	 p = p_0 - 0.5 \rho u^2
	\label{eq:pressure_calculation}
\end{equation}
}
\addedRIV{where $p$ represents the static pressure, $p_0$ is the total pressure, $\rho$ is the fluid density, and $u$ is the normal velocity.} \deletedRIV{Therefore, the lower and upper velocity limits are set according to \mbox{\cite{cosoli2013surface,notarstefano2008estimation,bolanos2014modelling}} where the surface velocities were estimated using a combination of data from radar, numerical models, and infrared satellite images of the Adriatic Sea. The authors noted that the surface currents are highly variable, with speeds ranging from less than 0.1 m/s to more than 0.5 m/s. The average surface velocities were around 0.1-0.2 m/s in most areas. Due to the high variability of surface currents, the boundary values produces velocities within the domain of up to 0.5 m/s, in order to accurately recreate surface velocities.}

Once the tangential velocities and total pressure values are set, the next step is to define velocity and pressure profile functions. This involves interpolating the tangential velocity and pressure values along the boundary length while ensuring that the values at the edges of the coastline are set to 0. This is then repeated in the optimization procedure until the boundary condition produces a velocity field that matches the velocity values of the drifter.

The turbulence is modeled using the \emph{k-$\omega$ shear stress transport} model \cite{ferziger2012computational}, which combines the strengths of the k-omega and k-epsilon turbulence models in order to enhance the accuracy and reliability of capturing complex turbulent flows. This hybrid approach divides the flow domain into two distinct regions: the near-wall region and the outer region. In the near-wall region, the model employs a wall-function approach, accurately capturing the near-wall turbulence behavior. In the outer region, the model acts as a free-stream model, providing reliable predictions of the turbulence characteristics away from the wall \cite{menter1992improved}. The turbulence variables ($k$, $\omega$) are computed using:
\begin{equation}
	k = \frac{3}{2}(\left|\mathbf{u}\right|I)^2
	\label{eq:turbulence_kinetic_energy}
\end{equation}
\begin{equation}
	\omega = \frac{k^{0.5}}{C_\mu^{0.25} L}
\end{equation}\\
where $k$ is turbulence kinetic energy,  $I$ is  turbulence intensity, $\omega$  is the specific dissipation rate, $C_\mu$ is a turbulence model constant and equals 0.09 while $L$ is the turbulent length scale.

\subsection{Numerical implementation}

The computational domain in considered test cases is either synthetic in nature or represents a realistic geographical region. For specific regions, the relevant data is obtained using Google Earth polygon extraction (available on \href{https://earth.google.com/web/@45.65501199,13.62242081,-4.16703122a,50265.47942331d,30y,0h,0t,0r/data=OgMKATA}{Google Earth}) and noted accordingly. Upon extracting polygons, the generation of an STL model becomes essential for the subsequent creation of a numerical mesh. A corresponding two-dimensional numerical mesh is then created using cfMesh \cite{juretic2015cfmesh} and implemented in an open-source CFD package OpenFOAM \cite{openfoam}. Given that equations \eqref{eq:ns_equation} and \eqref{eq:compressibility_equation} define a steady-state incompressible flow, simulations utilize the simpleFoam solver implemented within OpenFOAM. This solver employs a semi-implicit method for pressure-linked equations (SIMPLE) \cite{patankar1983calculation}. In terms of boundary conditions,  we treat the coastline as a solid wall by applying a no-slip (Dirichlet) boundary condition. When fluid flows out of the domain at a boundary face, the boundary condition for velocity is defined as the Neumann boundary condition, meaning that the velocity of the fluid at the boundary is extrapolated to the velocity inside the domain. When fluid flows into the domain, the open sea boundary switches to a Dirichlet boundary condition, where the velocity is calculated based on the flux in the patch-normal direction. Additionally, we defined tangential velocities because the flow entering the domain is not necessarily perfectly aligned with the inlet boundaries. Specifying a tangential velocity helps simulate more realistic inlet flow conditions, taking into account any swirl or tangential motion of the fluid. 

The pressure at the open sea boundary is defined using the Dirichlet boundary condition with a set range of values while the coastline is defined using the Neumann boundary condition. To ensure that the pressure value within the domain is well-defined, a reference cell is selected and assigned a pressure value of zero. This reference cell serves as a reference point for the pressure gradient calculations throughout the domain.

The \emph{k-$\omega$ shear stress transport} model is employed in our modelling approach, hence we defined turbulence values at the coastline using wall functions, while for the open sea, we specified the Neumann boundary conditions. Boundary conditions for all test cases are briefly summarized in Table \ref{tab:boundary_conditions}.

\begin{table*}[!htb]
	\footnotesize
	\centering
	\caption{An overview of the boundary conditions employed}
	\label{tab:boundary_conditions}
	\begin{tabularx}{\linewidth}{Xrrl} 
		\\
		
		\textbf{Field}							& Inlet/Outlet	& Coastline				\\
		\hline 	
		
		$u$										& pressureInletOutletVelocity	& noSlip    					\\
		
		$p$										& totalPressure			& zeroGradient						\\
		$k$										& fixedValue 					& kqRWallFunction	\\
		$\omega$								& fixedValue 			& omegaWallFunction			\\
		
		\\

	\end{tabularx}

\end{table*}

Second-order accuracy was predominantly employed in OpenFOAM simulations, with second-order gradient and Laplacian schemes. Notably, first-order schemes (Gauss upwind) were selectively used for divergence terms related to convective transport to enhance stability in regions of steep gradients. Time derivatives and interpolations were treated with default second-order and linear schemes, respectively, while the meshWave method calculated distances to the nearest wall. The described boundary conditions and overall setup were kept the same for all test cases. 

Detailed information about the numerical implementation, including figures of the numerical grid for each case, along with specifics regarding cell distribution, discretization, and modeling schemes, can be found on The Open Science Framework repository: \url{https://osf.io/pdtbh/}.

\section{Model fitting and optimization problem formulation}

The concept of model fitting implies an iterative procedure to determine velocity and pressure functions
at the domain boundaries. These functions define the boundary values that induce various types of flows within the domain. The objective of the model fitting process is to minimize the disparity between the results obtained using CFD and sparse measurements. Therefore, in order to formulate the optimization problem, functions need to be parameterized i.e. the optimization vector needs to be defined.

\subsection{Boundary condition parametrization}

In the proposed approach, the optimization algorithm iteratively improves the values from the optimization vector to minimize the error value, i.e. to minimize the difference between the simulated flow field and the reference flow. As a result, the accuracy and reliability of the flow field reconstruction heavily rely on the optimization vector. This vector encompasses the values of tangential velocity and pressure at the boundary control points:
\begin{equation}
	\mathbf{b} = \left(\mathbf{u}_{t,1}, p_1, \ldots,\mathbf{u}_{t,n_{CP}}, p_{n_{CP}} \right)^T
\end{equation}
where $n_{CP}$ is the number of boundary control points. 

The locations of boundary control points where velocity and pressure values are specified can be uniformly or non-uniformly distributed, and their placement is determined based on the length of the boundary. A negative velocity value in the optimization vector signifies that the fluid is leaving the domain, whereas a positive value indicates that it is entering the domain. The velocity at the edges of the coastline is 0, as the wall has a no-slip condition. Further, the pressure values enable the computation of flux, a critical factor in determining velocity in the patch-normal direction. Consequently, the velocity at the boundary control point is a combination of the tangential velocity set by the optimization algorithm and the velocity computed from the flux in the patch-normal direction. 

\begin{figure}[H]
	\centering
	\includegraphics[width=0.7\linewidth]{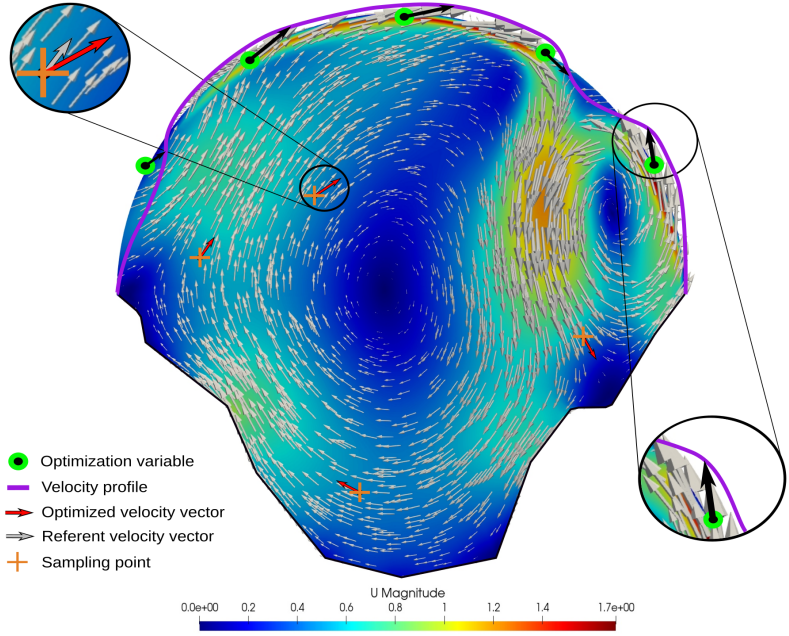}
	\caption{The figure illustrates the velocity profile generated using a specified optimization vector and how optimization variables (green dots) influence flow field reconstruction. The objective is to adjust the optimization variables until the red arrow (representing the current optimized velocity vector) aligns with the grey arrow (representing the reference velocity vector) for every sampling point (orange cross).  }
	\label{fig:optimized_veloctiy_field}
\end{figure}

To accurately represent realistic flows and account for the high variability of surface currents \addedRIV{in submesoscale domains}, the bounds for optimization variables are set to a range of -0.5 to 0.5 $m/s$ for tangential velocity and -0.05 to 0.05 $m^2/s^2$ for pressure at the boundary control points\addedRIV{, as discussed in subsection \ref{subsec:Stationary_2D_flow_model}}. These bounds are defined for numerical purposes, \replacedRIV{where the initial candidates in the optimization process are randomly distributed within this range as they do not significantly affect the qualitative outcomes. It is important to note that the resulting values for total pressure and tangential velocity at the boundary do not always match the assigned values, as they are adjusted in accordance with the solution of the Navier-Stokes (NS) equations in the internal domain. Additionally, using broader limits for optimization can be advantageous, as it promotes smoother convergence, despite increasing the size of the search space.}{specifically to ensure that simulations accurately produce realistic values of surface currents typically observed in submesoscale domains.} Once the values are assigned to the boundary control points, cubic spline interpolation is used to obtain values (velocity and pressure) for each cell along the length of the boundary. Subsequently, during the optimization process, as the optimization vector changes, the velocity profile is iteratively updated until the output (error value) matches the target fitness. Visual representation of velocity profile with referent and optimized velocity vectors can be seen in Figure \ref{fig:optimized_veloctiy_field} where a synthetic scenario called Simple bay case is created to showcase a workflow on a conventional domain with features outlined in Table \ref{tab:simple_bay_characteristics}. The simulation shows that certain parts of the domain, specifically the ones very close to the coastline, remain intact throughout the entire simulation. This means that the vortices do not reach or impact those areas.

\begin{table*}[!htb]
	\footnotesize
	\centering
	\caption{Characteristics of the Simple bay test case}
	\label{tab:simple_bay_characteristics}
	\begin{tabularx}{0.5\linewidth}{Xrl} 
		\\
		
		\textbf{Case characteristics}							& Simple bay 					\\
		\hline 	
		
		Test case type 												& Synthetic    					\\
		
		Domain area	[$km^2$]										& 24.6 							\\
		Number of boundaries 							& 1 						\\
		Total boundary length [$km$] 								& 9.4 						\\
		Coastline length [$km$]  								& 9.1 						\\
		Number of boundary control points 					& 5 					\\
		
		Max velocity in the domain [$m/s$]									& 0.2 					\\
		
		Presence of coastline/islands						& Coastline 					\\

		Number of cells 									& 4625 			\\
		Average cell size [$m$] 									& 73.02					\\
		\\

	\end{tabularx}
\end{table*}

\subsection{Objectives}\label{objectives}

A cost function is introduced to compute the difference between the reference values and calculated values. The reference values are discrete, point values, i.e. they are defined by their coordinates and corresponding velocity vector. During each evaluation, an OpenFOAM case is generated, where the entire velocity field is calculated. The velocity vectors at the coordinates of the sampling points representing the drifter locations are subsequently extracted. The cost function computes the drifter error $\epsilon_d$, which combines the error of the discrete points (drifters), using the equation:
\begin{equation}
	\epsilon_d(\mathbf{b})  = \dfrac{1}{n_{SP}} \sum_{i=1}^{n_{SP}} (\mathbf{u}_{r_i} - (\mathbf{u}_{s_i}(\mathbf{b})+\mathbf{u}_w^*))^2 
	\label{eq:drifter_error}
\end{equation}
where  $n_{SP}$ is the number of sampling points, $\mathbf{u}_{r_i}$ is the reference velocity vector, $\mathbf{u}_{s_i}$ is the simulation velocity vector at sampling point location $\mathbf{s}_i$ and $\mathbf{u}_w^*$ is wind velocity vector. 

The wind velocity vector is calculated through inner optimization for the entire field according to the:
\begin{equation}
	\begin{aligned}
		& \mathbf{u}_w^*(\mathbf{b}) =
		& & \underset{\mathbf{u}_w}{\text{argmin}}\left(\dfrac{1}{n_{SP}} \sum_{i=1}^{n_{SP}} (\mathbf{u}_{r_i} - (\mathbf{u}_{s_i}(\mathbf{b})+\mathbf{u}_w))^2\right)
	\end{aligned}
	\label{eq:wind_approx}
\end{equation}
where optimization tools from SciPy \cite{virtanen2020scipy} were used to find minimum values of $\mathbf{u}_w^*$. The computational time for inner optimization is negligible and has no important impact on the overall optimization process duration since it relies on a single surrogate model simulation parameterized with $\mathbf{b}$.

To assess the effectiveness of flow reconstruction, we introduced a field error variable, denoted as $\epsilon_f$, which represents the global error of the velocity field, similar to equation \ref{eq:drifter_error}.  The values of the velocity vectors at these field points are used to validate the field reconstruction and are not utilized in the optimization process:
\begin{equation}
	\epsilon_f(\mathbf{b})  = \dfrac{1}{n_{FP}} \sum_{j=1}^{n_{FP}} (\mathbf{u}_{r_j} - (\mathbf{u}_{s_i}(\mathbf{b})+\mathbf{u}_w^*))^2 
	\label{eq:field_error}
\end{equation}
where $n_{FP}$ is the number of field points, $\mathbf{u}_{r_j}$ is the reference velocity vector, and $\mathbf{u}_{s_j}$ is the simulation velocity vector at the field point location.

\replacedRIV{This approach, with the optimization objective $\epsilon_d$, and the observable $\epsilon_f$,  is consistently used in all test case scenarios.}{Given that $\epsilon_f$ for solely used as observation, the primary optimization objective $\epsilon_d$, and this approach is consistently employed across all test case scenarios.} In synthetic scenarios where the complete velocity field values are accessible, the number of field points can go up to the number of cells of the numerical grid. However, in real-case scenarios with a limited number of deployed drifters, there are no corresponding field points available to validate flow reconstruction. In such cases, the success of flow reconstruction relies solely on the data obtained at sampling points. If there is a substantial deployment of drifters (50 or more), or when sampling points represent HF-radar measurements (which consistently provide more than 50 data points), users have the flexibility to allocate a percentage for sampling and the rest for field points.

\subsection{Constraints}\label{constraints}

In order to attain viable solutions of the context of this simulation-based optimization approach, it is necessary to set appropriate constraints. In this approach, optimization constraints effectively correspond to simulation residuals, serving as a guide in the optimization process to achieve targeted fitness and numerically viable simulation results. If the residuals fall below their set limits, the constraints are considered satisfied. If not, a penalty is applied when the residuals do not meet the required criteria.

The pressure residual constraint, defined as:
\begin{equation}
	r_p(\mathbf{b}) \leq 1e^{-3}
	\label{eq:p_constraint}
\end{equation} helps to maintain consistent pressure values throughout the optimization, preventing pressure imbalances that could lead to unrealistic behavior. Velocity residual constraint\addedRIV{s for both velocity components restrict the velocity magnitude within a specified range, enabling controlled and physically feasible fluid motion. They are defined as:}\deletedRIV{, set as:}
\begin{equation}
	\addedRIV{r_{u_{x}}(\mathbf{b}) \leq 1e^{-4}}
	\label{eq:vel_x_constraint}
\end{equation} 
\begin{equation}
	\addedRIV{r_{u_{y}}(\mathbf{b}) \leq 1e^{-4}}
	\label{eq:vel_y_constraint}
\end{equation}

\deletedRIV{restricts the velocity magnitude within a specified range, enabling controlled and physically feasible fluid motion.} \emph{k} residual constraint, set as: 
\begin{equation}
	r_k(\mathbf{b}) \leq 1e^{-4}
	\label{eq:k_constraint}
\end{equation} ensures that the turbulent kinetic energy remains within acceptable bounds. The $\omega$ residual constraint, set as:

\begin{equation}
	r_\omega(\mathbf{b}) \leq 1e^{-4}
	\label{eq:omega_constraint}
\end{equation} restricts the specific dissipation rate of turbulence, ensuring its consistency with the underlying physics. The optimization process is collectively influenced by these constraints, ensuring the preservation of physical realism, stability, and accurate representation of fluid dynamics. These 5 constraints remain consistent across all test case scenarios.

\subsection{Multimodality}\label{multi_modal}

Initially, testing was performed on the widely studied cavity lid case problem. This scenario represents a situation in which a squared cavity (area of $1m^2$) is filled with fluid and the flow is induced by the movement of the lid defined with 3 control points. This motion triggers the formation of vortices and recirculation zones, which we aim to reconstruct. Velocities were assessed at 100 sampling points for 100 different sampling scenarios where tangential velocity at the inlet was set from -2$m/s$ to 2 $m/s$. \replacedRIV{The turbulence model was not enabled in this investigation as the only goal was to determine whether the velocity vector at a specific location could be produced by different boundary values. Given that Reynolds numbers were expected to reach up to 200, a steady-state and laminar model was employed.}{This resulted in Reynolds numbers of up to 200.} The structured mesh with 40,000 cells was chosen specifically to ensure high resolution and accuracy in capturing the complex dynamics of the cavity lid-driven flow. By employing \replacedRIV{these}{steady-state} simulations, we thoroughly explored how changes in Reynolds numbers affect the persistence, size, and dynamics of vortices and recirculation zones.

\begin{figure*}[!h]
	\centering
	\includegraphics[width=\linewidth]{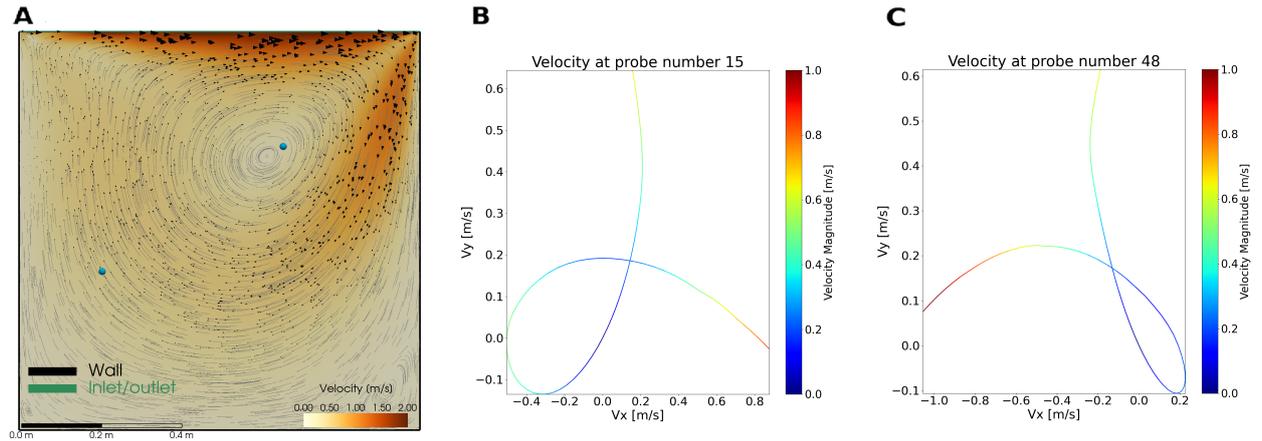}
	\caption{In total, 100 cavity lid cases are analyzed at 100 different sampling points within the domain to determine the multi-modality of the flow fitting problem. (\textbf{A}) Cavity lid case for $Re = 160$, where the cyan-coloured dots represent two observation locations of 100 chosen for the multimodal investigation. (\textbf{B})  shows the velocity variation at sampling location 15 (the left point on (\textbf{A})), with intersecting velocity curves showing that the same domain location can have the same velocity for different boundary conditions. (\textbf{C}) provides another example of intersecting velocity curves (the right point on (\textbf{A})). When examining all sampling points over the entire range, it was concluded that certain locations within the domain had the same velocities for different optimization vectors. This finding suggests a multimodal nature of the problem indicating that different optimization vectors can lead to identical results.}
	\label{fig:local_extremes}
\end{figure*}

Results indicated that a velocity vector at a chosen sampling point can \addedRIV{indeed} be induced by different combinations of boundary values. Consequently, achieving the desired reference velocity through optimization does not guarantee the accuracy of the flow field in its entirety. This observation is briefly described in Figure \ref{fig:local_extremes}. Based on this observation, local search methods are deemed unsuitable for our simulation-based optimization approach.

\section{Optimization methods and benchmark}

After detailing the optimization process, the identification of the most suitable optimization algorithm for the proposed flow field reconstruction approach is performed. In order to evaluate the performance of various methods, a predefined threshold is established, enabling us to assess efficiency, robustness, and scalability.

\subsection{Analysis and Classification of Optimization Outcomes}

The mean square difference of the velocities at sampling points, $\epsilon$\textsubscript{d}, is used as a fitness function in all assessed optimization tests i.e. the optimization is defined as follows:

\begin{equation}
	\begin{aligned}
		& \underset{\mathbf{b}}{\text{minimize}}
		& & \epsilon_d(\mathbf{b}) = \dfrac{1}{n_{SP}} \sum_{i=1}^{n_{SP}} (\mathbf{u}_{r_i} - (\mathbf{u}_{s_i}(\mathbf{b})+\mathbf{u}_w^*))^2\\
		& \text{subject to}
		& & \mathbf{b}_l \mathbf{\leq b} \leq \mathbf{b}_u
	\end{aligned}
\end{equation}

Convergence is considered achieved when the drifter error threshold $\epsilon$\textsubscript{d} =0.0001 is reached, which effectively represents a drifter velocity error in $m/s$. The flow field reconstruction error is additionally calculated for all optimization tests \addedRIV{with the field error threshold set to $\epsilon$\textsubscript{f} =0.0003} \deletedRIV{The field error threshold is set to $\epsilon$\textsubscript{f} =0.0003}, which effectively represents a field velocity error in $m/s$. \replacedRIV{Based on conducted tests, we determined that setting the field error threshold to be three times greater than the drifter error ensures an accurate representation of the reconstructed flow.}{$\epsilon$\textsubscript{f} is only monitored and planned to be three times greater than the drifter error, maintaining a proportional difference for an accurate representation of the reconstructed flow.} All results below these thresholds are deemed adequate.

\begin{figure}[!htb]
	\centering
	\includegraphics[width=0.7\linewidth]{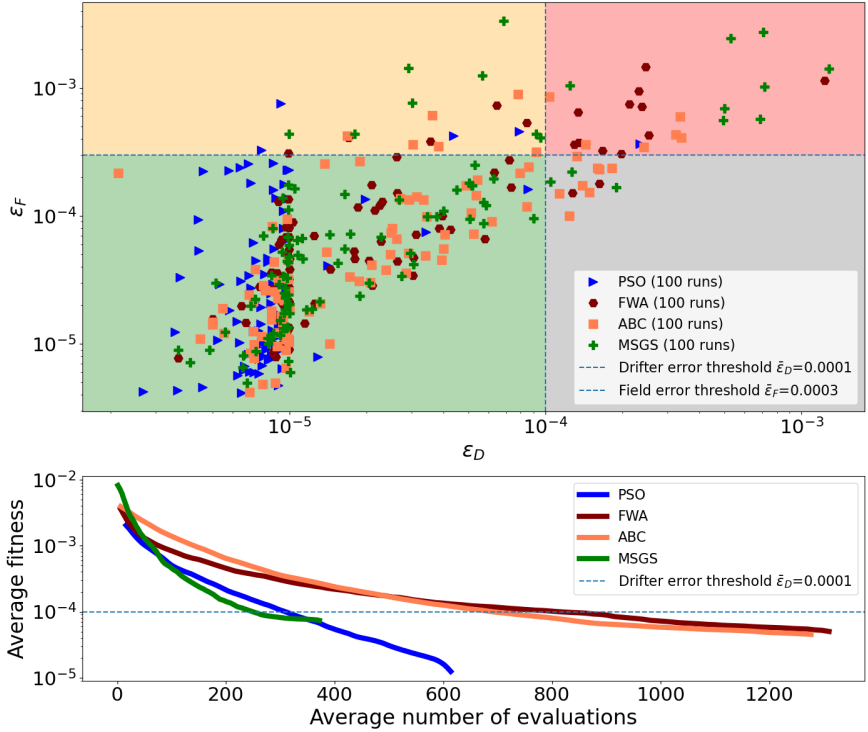}
	\caption{Algorithms used for numerical optimization and their applicability for flow field reconstruction. The four colored squares represent four different result groups mentioned in the above text, with 6.5\% of the results falling in the red area, 5.75\% in the orange area, 3.25\% in the grey area,  and the majority, 84.5\%, in the green area. The lower subplot shows the average convergence across all optimizations for each optimization algorithm. It is evident that MSGS, a local search algorithm, converges first. However, its overall fitness is on average worse than that of global search algorithms. Furthermore, compared to PSO, it often does not reach the specified threshold. PSO achieves an excellent balance between performance (second only to MSGS) and accuracy. This can be clearly seen from the narrow grouping (blue triangles) of results around $\epsilon_d$ = 1e-5. ABC and FWA demonstrate comparable fitness to MSGS, albeit with an extended optimization time.}
	\label{fig:investigation}
\end{figure}

Based on values computed for $\epsilon$\textsubscript{d} and $\epsilon$\textsubscript{f} , all optimization
results can be classified into 4 distinct groups which are shown in Figure \ref{fig:investigation}:

\begin{itemize}
	\item $\epsilon_d$ > 0, $\epsilon_f$ > 0 (red area) 
	
	A scenario where optimization failed to achieve the desired thresholds for both drifter and field error.
	\item $\epsilon_d$ > 0, $\epsilon_f$ $\approx$ 0 (gray area)
	
	An uncommon scenario where optimization successfully reaches the target threshold for field error but falls short in achieving the same for drifter error.
	\item $\epsilon_d$ $\approx$ 0, $\epsilon_f$ > 0 (orange area)
	
	A scenario where optimization effectively achieves the target threshold for drifter error, but the reconstructed field deviates from the reference field.
	\item $\epsilon_d$ $\approx$ 0, $\epsilon_f$ $\approx$ 0 (green area)
	
	The preferred scenario in which optimization successfully achieves target thresholds for both drifter and field error, signifying a successful reconstruction of the surface flow.
\end{itemize}

In order to determine the appropriate optimization algorithm for the proposed flow field reconstruction approach, evaluation of the methods provided by the Python numerical optimization module Indago  \cite{indago} was performed. During the initial testing, several additional methods were evaluated, but we opted to present the most successful ones. Our aim was to determine the global search methods that best fit our modeling requirements, as we excluded local search methods due to the presence of multimodality explained in \ref{multi_modal}. Particle Swarm Optimization (PSO) was initially selected due to its widespread application and proven effectiveness in efficiently optimizing complex objective functions \cite{kennedy1995particle}. The Fireworks Algorithm (FWA) was chosen as it could offer improvements over PSO, particularly in terms of convergence speed and global exploration capabilities \cite{tan2010fireworks}. The Artificial Bee Colony (ABC) algorithm was included for its robust performance in handling complex search spaces across various optimization scenarios \cite{karaboga2009comparative}. Furthermore, to showcase the distinctions in optimization results between local and global search techniques, the MSGS algorithm \cite{indago}, a modification of the GPS-MADS method \cite{audet2006mesh}, was chosen to represent local search methods, offering a robust approach to optimization in challenging scenarios where derivative information is unavailable. PSO demonstrates proficiency in navigating expansive solution spaces, FWA excels in scenarios with multiple optima, and ABC proves effective in addressing constrained optimization problems, all well suited for our modeling concept \cite{mohammadi2023intelligent}.

For this investigation, ten \replacedRIV{distinct}{different} Simple Bay test cases were \deletedRIV{randomly} generated, \replacedRIV{each with a randomly assigned boundary conditions.}{each serving as a reference case.} \replacedRIV{The optimization algorithms were tested 10 times for each reference case, with each test using random initialization of optimization vector and 10 randomly chosen sampling points for flow reconstruction. This approach ensured that the algorithm's performance was evaluated under diverse conditions resulting}{Each optimization algorithm was tested 10 times on each reference case, using 10 sampling points to reconstruct the flow field. This resulted} in a total of 100 optimizations performed by each algorithm. Since this is a synthetic scenario, wherein the case is predetermined and wind omitted, $\mathbf{u}_w^*$ is treated as 0. The results are shown in Figure \ref{fig:investigation}.

It is evident that, on average, all methods successfully met the specified threshold, albeit at different speeds. The target objective threshold was intentionally reduced to 1e-5 in order to improve the accuracy of the reconstructed field. Still, we retained a successful reconstruction threshold of 1e-4 for the optimization process, as indicated in Figure \ref{fig:investigation}. Notably, the local search technique MSGS demonstrated impressive speed, albeit with higher fitness than global search methods. Additionally, MSGS can be trapped in local optima, as we have previously underscored in the context of the cavity lid scenario. Therefore, we opted for global search methods in our modeling approach, with Particle Swarm Optimization (PSO) emerging as the most effective global search method. Although effective, PSO is slower than the considered local search method. \replacedRIV{It is important to note that the results are case-dependent, as each test case involves a different number of control points, which increases the problem’s dimensionality, and varying measurement locations that impact the optimization process. Nonetheless, we expected similar results across different test cases due to the synthetic nature of the simulations, despite differences in data sources. Considering the variations in size and complexity among our test cases, we have chosen PSO as the preferred method for our modeling approach.}{Given the variation in size and complexity among our test cases, we have selected PSO as the preferred method for our modeling approach.} This choice enables us to accurately capture more complex flows without becoming stuck in local extremes. The approach was also tested using a total of 50 optimization runs per algorithm and the differences were below 2\%. The results were in accordance with presented observations and confirm drawn conclusions.

\subsection{Reducing simulation time with model field initialization}

The optimization process aims to achieve the best solution through iterative modification of the optimization vector. This implies that every case starts from internal field values that are equal to 0, but the boundary conditions vary. Occasionally, specific combinations of optimization vector values result in simulations that either fail to converge or require significant time to converge, thus extending the overall optimization time. In order to reduce simulation time and improve optimization efficiency a field initialization approach is introduced where the internal field values obtained from the currently best solution are set as the initial state for new simulations. The rationale behind this stems from the observation that, during optimization, candidates tend to converge toward the best solution, leading to a highly similar flow. Consequently, initializing a simulation to the resulting state of that similar flow allows for faster simulations, i.e. enables the simulation to converge in fewer iterations and therefore achieve the results more quickly. By initializing the simulations with values from the best results, simulation times can be reduced by up to 20\%.

\begin{figure}[!h]
	\centering
	\includegraphics[width=0.7\linewidth]{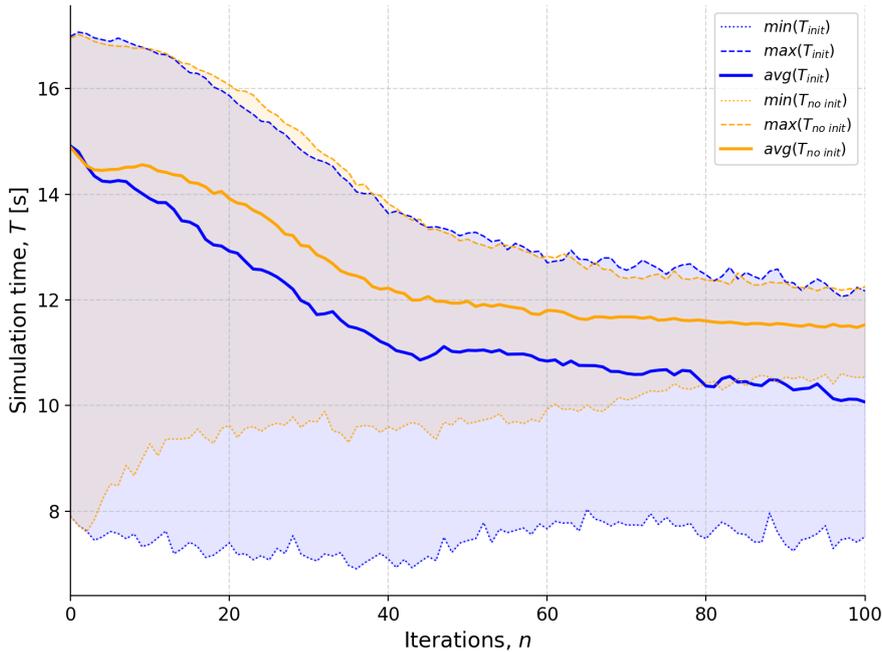}
	\caption{Computational improvements achieved through the use of simulation initialization. Colored segments illustrate simulation run times, which are bounded by the average maximum and minimum times for each optimization iteration. Dotted lines denote the average minimum simulation times, while dashed lines represent the average maximum times, indicating that all simulation run times are expected to fall within this range. The cumulative impact and the overall reduction of the simulation time are evident. Initially, the improvement is negligible, but after 100 iterations can reach up to 20\%.
	}
	\label{fig:speed_up}
\end{figure}

The impacts of this improvement may not be immediately evident in smaller simulations or domains where convergence is easily achieved. In fact, it can even initially prolong the optimization time in some cases. However, in larger domains characterized by complex flows and more demanding convergence requirements, the benefits are significant. \addedRIV{The results of the model field initialization approach for the Simple Bay test case are shown on Figure \ref{fig:speed_up} where the comparison is conducted based on 300 pairs of optimizations using both initialization and no-initialization approaches. For accurate comparison, each pair used the same initial conditions, the same flow to reconstruct, and the same sampling points obtained through random sampling, with the optimization seed remaining consistent for both approaches. It is evident that the field initialization contributes to the overall reduction in computational time. This approach has been tested with other test cases and similar behavior was noted.}

\section{Assessment of flow reconstruction accuracy depending on the \protect\addedRIV{effective} number of sampling points}

The accuracy of the flow field reconstruction depends not only on the number of sampling points but also on the location of these points within the domain. A larger number of sampling points means more information for the optimization algorithm and therefore, typically, stable convergence and accurate results. However, if the locations of the sampling points are very close to each other, they effectively provide the same information about the velocity field, hence the number of effective sampling points is lower. To determine how many sampling points provide useful information, an effective number of sampling points {$\eta$} is introduced.

For each sampling point, we decided to estimate the area of influence using the two-dimensional Gaussian function. The impact of sampling point $\mathbf{s}_i$ to its surroundings is characterized with
\begin{equation}
	\phi_i(\mathbf{x}) = \frac{1}{2\pi\sigma^2}\exp
	\left(
	-\frac{\left({\mathbf{x} }-{\mathbf{s}_i}\right)\cdot\left({\mathbf {x} }-{\mathbf{s}_i}\right)}{2\sigma^2}
	\right)
	\label{eq:gaussian_function}
\end{equation} where $\sigma$ is standard deviation and $(\cdot)$ is a dot product. The scaling $\frac{1}{2\pi\sigma^2}$ ensures the volume under the Gaussian function is always equal to 1, regardless of $\sigma$.

Our goal is to cover the domain with $n_{SP}$ sampling points where, in the ideal case, each covers the circle of a radius equal to three standard deviations (99.7\% of the volume under $\phi$ is within three standard deviations from the sampling location $\mathbf{s}$). Considering the influence of all sampling points is equal, i.e. $\sigma$ is the same for all $i$, we can determine $\sigma$ from
\begin{equation}
	9\sigma^2 \cdot \pi \cdot n_{SP} = |\Omega| 
	\label{eq:sigma}
\end{equation}
where $|\Omega|$ is the area of the domain $\Omega$.

\begin{figure}[H]
	\centering
	\includegraphics[width=0.7\linewidth]{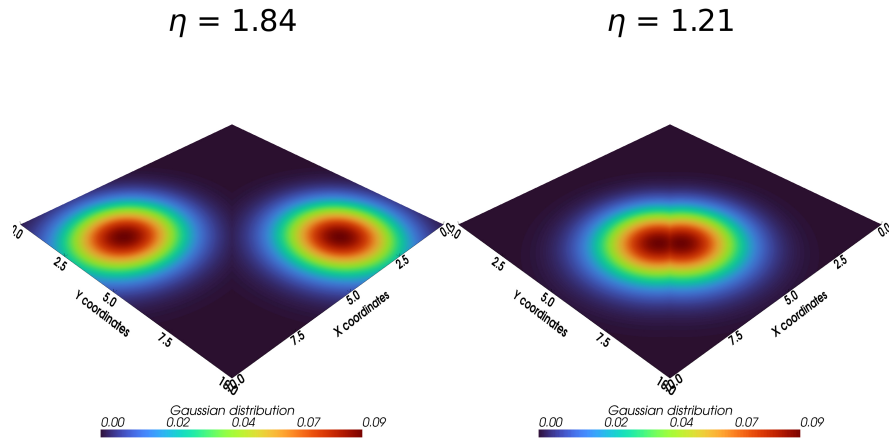}
	\caption{Illustration of the effective number of sampling points. The sampling points on the left are distant enough and hence are able to provide unique information about different parts of the domain. In this case, \emph{$\eta$} is smaller than \emph{n} because the domain is rectangular and the corners of the domain cannot be reached using the Gaussian distribution. The sampling points on the right are close to each other and, as shown, zones overlap, which means that sampling points partially provide the	same information and therefore \emph{$\eta$} is much smaller than \emph{n}.}
	\label{fig:gaussian_distribution}
\end{figure}

From \eqref{eq:sigma} $\sigma$ can be calculated explicitly.With the locations of sampling points and standard deviation $\sigma$ known, we can calculate Gaussian function $\phi_g(x,y)$ around each sampling point in the domain. Consequently, if these points are close to each other, their influence will overlap. In order to calculate an effective influence, a maximum value of $\phi$ is taken. Finally, an  effective number of sampling points {$\eta$} is calculated as:
\begin{equation}
	\eta = \int_{\Omega} \max(\phi_1,...,\phi_{n_{SP}})\,d{\mathbf{x}}
	\label{eq:effective_number_of_drifters}
\end{equation}
where it is expected for resulting $\eta$ to be a positive real number and to be less than the number of sampling points ($0 < \eta \leq n_{SP}$, for ideally distributed sampling points $\eta=n_{SP}$). 
Visual representation of \emph{$\eta$}  can be seen in Figure \ref{fig:gaussian_distribution}.

\begin{figure}[H]
	\centering
	\includegraphics[width=0.7\linewidth]{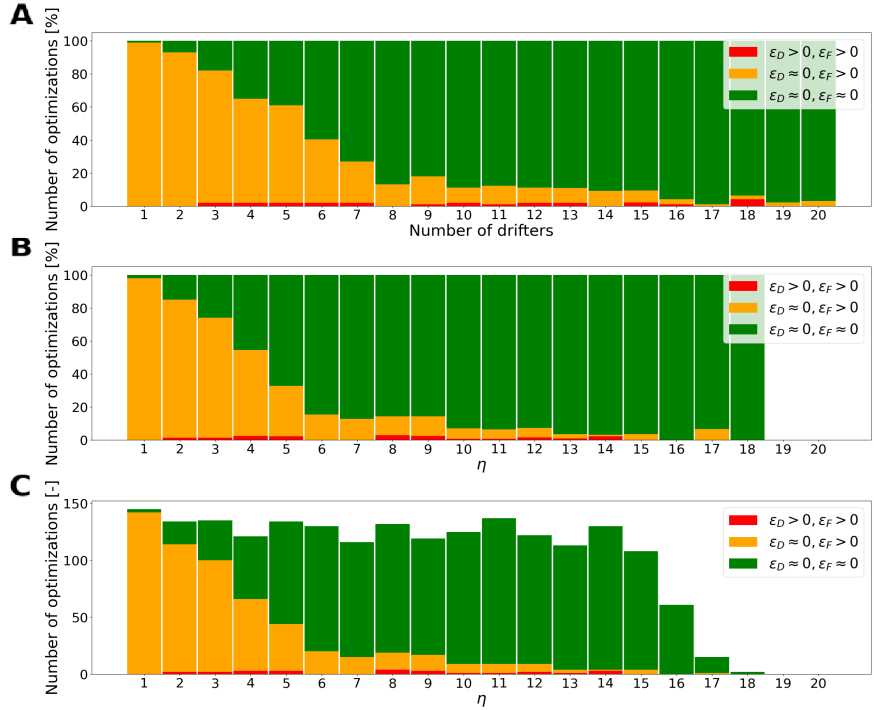}
	\caption{Results for 100 optimization runs for 1-20 sampling points randomly distributed across the domain. (\textbf{A}) As expected, the results are not reliable with a small number of sampling points (e.g. 1-3), but as the number of sampling points increases, this behavior changes. This becomes even clearer in the (\textbf{B}), which shows relative distribution of optimization success based on the effective number of sampling points. (\textbf{C}) shows the absolute distribution of optimization success over the different effective sampling point numbers, highlighting the correlation between sampling points and optimization outcomes.}
	\label{fig:Drifters_accuracy_with_eta}
\end{figure}

The impact of $\eta$ on the flow reconstruction results is shown in Figure \ref{fig:Drifters_accuracy_with_eta}. A range of potential scenarios resulting from 100 optimizations for each configuration of 1-20 sampling points randomly positioned inside the Simple bay domain is considered. When the number of sampling points is small (e.g. 1-3), the drifter error threshold is reached, but the $\epsilon$\textsubscript{f} is significant (orange). As the number of sampling points increases, more data is available to the optimization algorithm, which leads to a better field reconstruction (green). The orange area does not represent an unsuccessful optimization, but rather a case where the number of sampling points is not sufficient to fully reconstruct the flow. This can be visible from the second plot where the orange area disappears completely, as more unique data is available to reconstruct the entire flow. \replacedRIV{While the second plot illustrates the relative distribution of optimization success, the third plot shows the absolute count of optimization runs for specific values of $\eta$ highlighting the very small number of runs associated with higher values of $\eta$. Given that realistic flow fields can move drifters to various positions, we simulated different drifter locations using a random distribution of sampling points to evaluate how these positions impact our reconstruction method. As a result of the random sampling approach, there are no data points for $\eta$ = 19 and  $\eta$ = 20 due to inadequate domain coverage stemming from employed random sampling. Overall, the observed expansion of the green area indicates that the accuracy of the flow field reconstruction improves as the number of sampling points increases.}{The observed trend of the expansion of the green area suggests that the reconstruction of the flow field becomes more accurate as the number of sampling points increases.}

\begin{figure*}[!h]
	\centering
	\includegraphics[width=\linewidth]{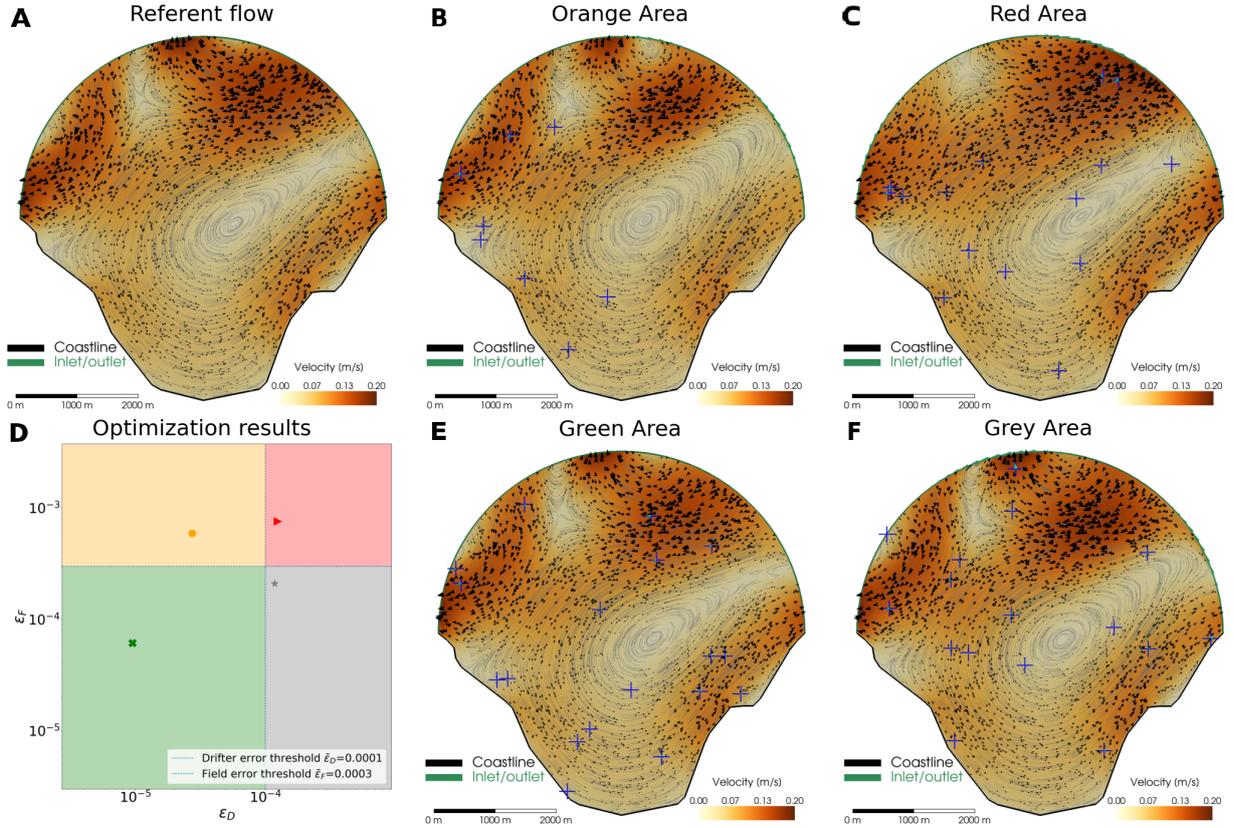}
	\caption{(\textbf{A}) The reference flow we aim to reconstruct. (\textbf{B}) With 8 sampling points, it is possible to accurately calculate the velocities at the sampling points locations, but the desired threshold for the entire field is not achieved. (\textbf{C}) 15 sampling points were used in this scenario, but only the main vortex is captured, with the flow in the northern part of the domain deviating significantly. (\textbf{D}) Displays the placement of optimization results within the scenario area. (\textbf{E}) Represents the green area scenario with 18 sampling points, where both drifter error and field error satisfy the desired threshold, conclusively indicating the successful reconstruction of the flow. (\textbf{F}) Case with 16 sampling points where the desired threshold for field error is satisfied but not for the drifter error.}
	\label{fig:simple_bay_opt_results}
\end{figure*}

Figure \ref{fig:simple_bay_opt_results} summarizes possible scenarios of the optimization process, with a visual comparison of the different scenarios and the reference flow. The reference flow we aim to reconstruct has two dominant flows in the northern and western parts of the domain, with a main vortex and a circular movement near the coastline. Reference flow can be reconstructed in different ways depending on the placement of the sampling points and the threshold used to determine if the reconstruction is successful. The figure confirms our hypothesis that the orange scenario ($\epsilon_d$ $\approx$ 0, $\epsilon_f$ > 0) as the number of drifters increases will eventually transition into the green area ($\epsilon_d$ $\approx$ 0, $\epsilon_f$ $\approx$ 0). On the other hand, the red scenario ($\epsilon_d$ > 0, $\epsilon_f$ > 0) corresponds to an insufficiently resolved problem where increasing the number of drifters may not improve the solution because the optimization algorithm did not have enough information in key regions. The grey scenario($\epsilon_d$ > 0, $\epsilon_f$ $\approx$ 0) provides a flow field that is similar to the reference flow, with slight variations in the velocities of the sampling points. These are rare cases, attributable to numerical errors, and are consistently near the threshold. While this area bears great similarity to the reference case, the measurements at the sampling points locations differ just enough to fall short of the prescribed threshold. These scenarios confirm that the placement of the drifters plays an important role; greater uniformity in their distribution could enhance the accuracy of flow reconstruction. Although adjusting the thresholds for drifter error and field error may yield somewhat different results, the findings suggest that the thresholds depicted in the figure are appropriate for flow reconstruction, given that the absolute error remains below 0.03 cm/s.

\section{Validation}

To validate our simulation-based optimization approach, we created three test cases in addition to the Simple bay test case. Each of these test cases possesses distinct features that set it apart from the others. The unique characteristics of each test case are detailed in Table \ref{tab:case_characteristics}.

\begin{table*}[!htb]
	\footnotesize
	\centering
	\caption{Characteristics of the validation test cases}
	\label{tab:case_characteristics}
	\begin{tabularx}{\linewidth}{Xrrrl} 
		\\
		
		\textbf{Case characteristics}							 			& Open water	& Gulf of Trieste	& Vis					\\
		\hline 	
		
		Test case type 												    		& Synthetic 		& Realistic  	& Realistic			\\
		
		Domain area [$km^2$]										 			& 28.3 		& 498.94			& 2273.9					\\
		Number of boundaries 							 			& 1 		& 1			& 5						\\
		Total boundary length [$km$]										& 125.65	& 20.52			& 120.97					\\
		Coastline length [$km$]								 			& -			& 86.64			& 197.61					\\
		Number of boundary control points 							& 12		& 5			& 14			& -					\\
		
		Max velocity in the domain [$m/s$]									 			& 0.3			& 0.2		& \protect\replacedRIV{0.35}{0.3.5}						\\
		
		Presence of coastline/islands						 	& No coastline			& Coastline			& Coastline and islands						\\

		Number of cells 										& 9600					& 8262			& 12856				\\
		Average cell size [$m$]										& 361.75				& 245.74		& 412.41			\\
		\\

	\end{tabularx}

\end{table*}

For each test case, different optimization requirements and parameters (e.g. bounds and variables) are set. A comprehensive overview of optimization parameters and characteristics is presented in Table \ref{tab:optimization_characteristics}.

\begin{table*}[!htb]
	\footnotesize
	\centering
	\caption{Optimization parameters for the validation test cases}
	\label{tab:optimization_characteristics}
	\begin{tabularx}{\linewidth}{Xrrrrrrl} 
		\\
		
		\textbf{Optimization characteristics}							 	& Open water	& \multicolumn{2}{r}{Gulf of Trieste}		& Vis			\\
		&				& Drifters					& HF-radars							& HF-radars	\\
		\hline 	
		
		Number of sampling points $n_{SP}$																& 35 					& 14						& 225								& 555					\\
		
		Number of objectives										 			& 2						& 2							& 2									& 2					\\
		
		Number of constraints 														& 6						& 6							& 6									& 6							\\
		
		Number of optimization variables 										& 20					& 6							& 6									& 28					\\
		
		Tangential velocity (bounds) [$m^2/s^2$]				 			& -0.2 - 0.2					& -0.5 - 0.5							& -0.5 - 0.5			& -0.5 - 0.5				\\
		
		Pressure (bounds) [$m^2/s^2$]								 			& -0.025 - 0.025					& -0.05 - 0.05							& -0.05 - 0.05			& -0.05 - 0.05				\\
		
		Average simulation time [$s$]  										    		& 8.13 					& 28.51  					& 32.28 							& 32.03					\\
		
		Average optimization time [$min$]									& 90.84					& 135.727					& 133.54							& 110.89				\\

		\\

	\end{tabularx}

\end{table*}

\subsection{Open water}

\begin{figure*}[!h]
	\centering
	\includegraphics[width=\linewidth]{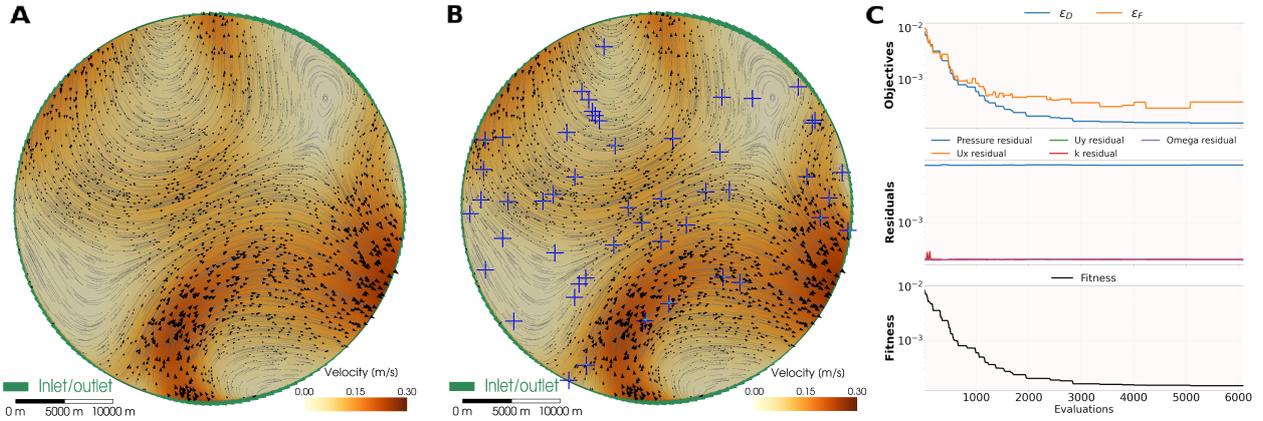}
	\caption{The figure depicts an Open water case, where the diameter of the domain is 40 km. Since the domain is very large, a total of 30 sampling points are used to reconstruct the flow. (\textbf{A}) The reference flow contains two main vortices located in the northern and southern parts of the domain. The maximum velocity of 0.3 m/s is in line with expectations for a domain of this size. (\textbf{B}) The reconstructed flow with 30 sampling points, is shown as blue crosses. On visual inspection of the field, the similarity to the reference flow is evident, with all vortices and movements accurately captured. (\textbf{C}) Insight into the inlet/outlet boundary with 10 control points leading to 20 optimization variables. The first subplot shows the alignment of field error and drifter error, whereby consistency is maintained throughout the entire optimization process. The second subplot shows that the pressure residual is at its highest, but successfully follows the limits defined by the optimization algorithm. The fitness plot shows an initial steep drop within the first 1000 evaluations, followed by a gradual descent requiring approximately 4000 evaluations to achieve the target fitness. }
	\label{fig:open_water_opt_results}
\end{figure*}

The Open water test case shown in Figure \ref{fig:open_water_opt_results} represents a particular type of problem where the coastline is missing. This is a common problem when modeling and reconstructing e.g. ocean flow field. The entire boundary can vary, which makes the optimization process challenging. Despite these optimization challenges, this case is of great interest, applicable to small bays, and potentially even larger sea regions. Given the considerable size of this domain(20 $km$ radius), 30 sampling points were positioned within the domain to reconstruct the surface flow\addedRIV{, resulting in an effective number of sampling points \emph{$\eta$} = 23.12}. Like the Simple bay case, this scenario is synthetic, with $\mathbf{u}_w^*$ being 0. \replacedRIV{Figure \ref{fig:open_water_opt_results} shows that the reconstructed flow exhibits characteristics similar to the reference flow. However, achieving this green area scenario required over 4000 evaluations through the optimization process.}{It can be seen from Figure \ref{fig:open_water_opt_results} that the reconstructed flow has similar characteristics to the reference flow, although 4000 evaluations were required to achieve this through the optimization process.}

\subsection{Gulf of Trieste}

The Gulf of Trieste, situated in the northernmost part of the Adriatic Sea, is a very shallow bay with an area of more than 500 square kilometers. Surface velocities in the Gulf of Trieste have been extensively studied by\cite{cosoli2013surface}, who utilized a combination of high-resolution simulations from the Northern Adriatic Princeton Ocean Model (NAPOM), HF-radar data, and acoustic Doppler current profilers (ADCPs) to estimate surface currents. Their findings indicate significant variability in surface currents, with velocities ranging from less than 0.1 m/s to over 0.5 m/s. Notably, the root-mean-square-error (RMSE) differences between radar and model estimates fall predominantly within the range of 8.6 to 11.2 cm/s for 80\% of the data points.

Complementing this work, \cite{bellomo2015toward} conducted a detailed validation of HF-radar velocity data against direct measurements of Lagrangian velocity using CODE drifters. Their analysis revealed an RMSE difference of about 10 cm/s for radial velocities derived from HF-radar signals, where a value within the 5–15 cm/s range is typically accepted for such measurements \cite{chapman1997accuracy, ohlmann2007interpretation, molcard2009comparison, enrile2018evaluation}. This consistency with previous observations in the Gulf of Trieste further underscores the reliability of these methods.

Similarly, \cite{bolanos2014modelling} utilized the MIKE3/21 modeling system, along with measurements of wind, waves, currents, and water levels at a single location, to investigate the current dynamics of the northern Adriatic basin. Their study evaluated model sensitivity to different parameterizations and implementation strategies, comparing results with available in-situ observations (waves, currents, surface elevation, and water temperature) and a high-resolution modeling system (COAWST) implemented in the same area. Over a one-year simulation period, validation against surface elevation, wind, and wave data indicated strong model performance, comparable to the COAWST implementation, with surface current speed RMSE errors from observations at the “Acqua Alta” platform being 13 cm/s.

\begin{figure*}[!h]
	\centering
	\includegraphics[width=\linewidth]{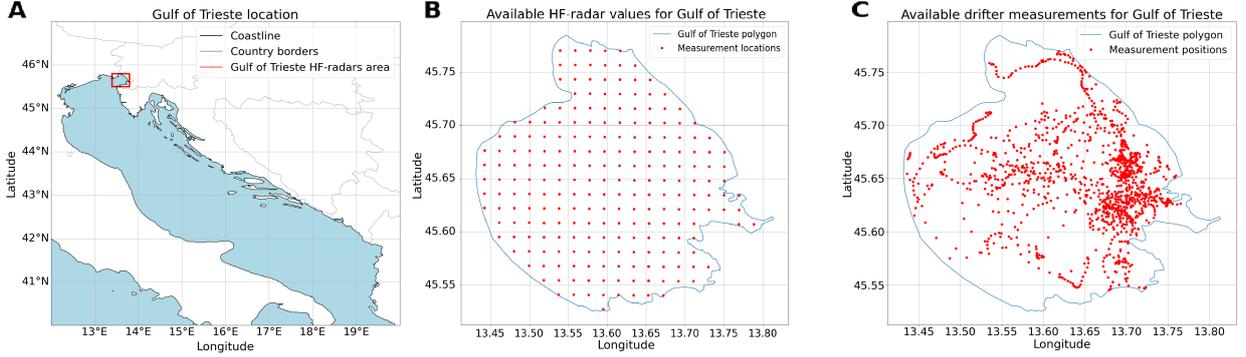}
	\caption{Figures show the Gulf of Trieste location in the Adriatic Sea and locations of available data based on drifter deployment and HF-radars.  (\textbf{A}) Gulf of Trieste in the northern part of the Adriatic Sea where its area is approximately 550 square kilometers (\textbf{B}) 225 available HF-radar data locations within the domain of the Gulf of Trieste for April 2012, (\textbf{C}) Location of 44 drifters released within the domain of the Gulf of Trieste from 23rd April 2012 to 4th May 2012.}
	\label{fig:trieste_data}
\end{figure*}

These studies collectively highlight the dynamic nature of the Gulf's surface currents, influenced by a complex interplay of local and regional factors. An overview of the Gulf of Trieste, including the locations of available measurement sites, is provided in Figure \ref{fig:trieste_data}. This figure illustrates the spatial distribution of the observational instruments, offering a comprehensive understanding of the measurement network employed to capture the Gulf's oceanographic conditions.

\begin{figure*}[!h]
	\centering
	\includegraphics[width=\linewidth]{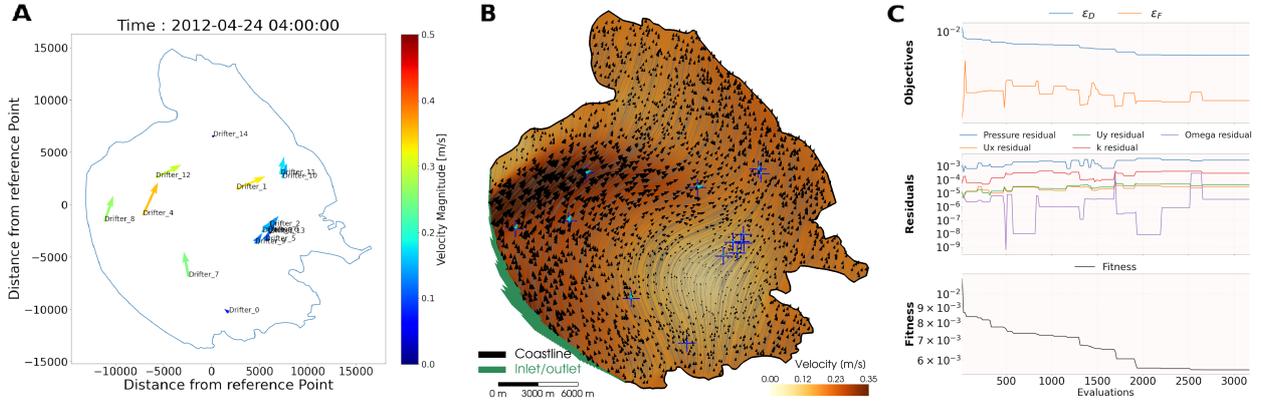}
	\caption{Figures show surface flow reconstruction results based on drifter measurements at 4 am on April 24th, 2012: (\textbf{A}) Displays the 15 available drifter locations and the corresponding velocity magnitudes at the specified time. The velocity magnitudes observed are below 0.4 m/s. (\textbf{B}) Illustrates the reconstructed surface flow from measurements obtained from the 15 available drifters, incorporating wind velocity where $\mathbf{u}_w^* = [5.88e^{-2},1.72e^{-1}]$ \protect\addedRIV{m/s}, to capture the actual flow dynamics. (\textbf{C}) This plot presents the changes in velocities error and residuals throughout the optimization process. It is observed that omega residual oscillates more than other parameters, but remains within the allowed range. The fitness plot indicates that the optimization did not reach the set threshold. However, the obtained results, including the visual representation and RMSE, which is 7.7 cm/s for this particular case, are considered highly satisfying.}
	\label{fig:trieste_drifters}
\end{figure*}

The validation of the Gulf of Trieste surface flow was carried out using data obtained as part of the TOSCA experiment in April 2012 \cite{magaldi_marcello_g_2019_3245409}. There are no more than 20 drifter locations providing velocity measurements. To perform flow reconstruction based on drifter and HF-radar measurements, we used the same specified time to assess the impact of the measurement data. Drifter measurements taken at 4 am on April 24th, 2012, reveal 15 available drifter locations that are closely positioned, resulting in an effective number of sampling points, \emph{$\eta$}, of 9.59, as depicted in Figure \ref{fig:trieste_drifters}. Since only 15 drifters were available for this specific time, one drifter was utilized to track field error, resulting in a noticeable gap between drifter error and field error in the subplot (\textbf{C}). For this realistic case, only three control points were placed, equating to six optimization variables. This decision was made due to the absence of vortices in the flow, with a dominant flow observed from the southern to the northern part of the domain.

\begin{figure*}[!h]
	\centering
	\includegraphics[width=\linewidth]{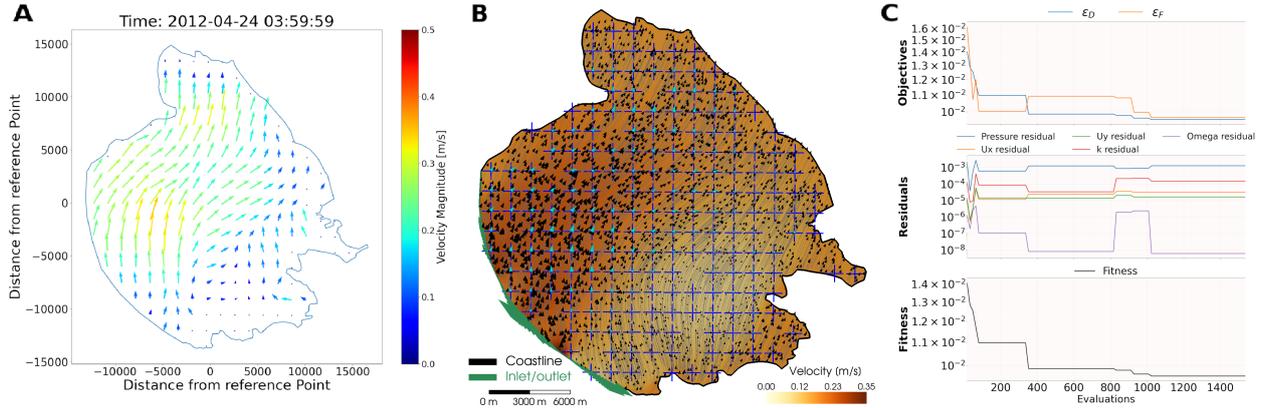}
	\caption{Figures shows surface flow reconstruction results based on HF-radar measurements at 4 am on April 24th, 2012: (\textbf{A}) Figure displays the 225 HF-radar data locations, showcasing the corresponding velocity magnitudes at the specified time. (\textbf{B}) The reconstructed surface flow field for a case with 202 sampling points and 23 field points (10\% reduction). Employed wind velocity is $\mathbf{u}_w^* = [6.37e^{-2},1.45e^{-1}]$ \protect\addedRIV{m/s}. Cyan arrows represent optimized velocity vectors that are aligned with HF-radar measurements, providing a visual representation of the accuracy of the reconstruction. (\textbf{C}) presents oscillations in omega residual for the first 1000 evaluations, which then stabilize towards the end of the optimization. The fitness plot indicates a steep decline in the initial 400 evaluations, followed by fine-tuning until the completion of the optimization process. While the optimization did not achieve the desired threshold, the results are considered satisfying, especially when visually comparing the reference flow to the optimized one. Numerically, RMSE of 10.1 cm/s is achieved.}
	\label{fig:trieste_radars}
\end{figure*}

An analysis of the HF-radar data reveals that the flow vectors do not obey the principle of conservation of mass. This is evident from the fact that the velocity vectors are directed towards the coastline, as illustrated in Figure \ref{fig:trieste_radars}. This deviation can be attributed to the significant influence of the wind on the surface flow, resulting in directional changes that steer the vectors toward the coastline. Previous authors \cite{cosoli2013surface, querin2021multi} themselves acknowledged that local conditions have a significant influence on surface velocities, which prompted us to include wind velocity in our evaluation function. The evaluation function extracts the velocities at the drifter locations and simultaneously integrates a wind velocity into the velocity vectors spanning the entire domain in x and y directions. As a result, the mass flux remains consistent, but the vectors tend to align directly with the coastline, providing a more realistic representation of the data. Following a similar approach as demonstrated in Figure \ref{fig:trieste_drifters}, only three control points were placed, resulting in six optimization variables, as reflected in the first subplot. Unlike the drifter scenario, in this case, significantly more measurement data is available. Therefore, we tested the accuracy of our method by reducing the data by 10\%, 50\%, and 90\%, with HF-radar measurements employed to track field error. An overview of conducted assessments for the Gulf of Trieste case is given in Table \ref{tab:trieste_case_characteristics}. Presented values are RMSE averages for 50 optimization runs.

\begin{table*}[!htb]
	\footnotesize
	\centering
	\caption{Data availability for the Gulf of Trieste case}
	\label{tab:trieste_case_characteristics}
	\begin{tabularx}{\linewidth}{l@{\hspace{2cm}}rrrrrr}
		\textbf{Data Type} & Available data points & Reduction [\%] &  $n_{SP}$ &  $n_{FP}$ & $\overline{RMSE}_d$ [cm/s] & $\overline{RMSE}_f$ [cm/s] \\
		\hline
		\multirow{2}{*}{\textbf{Trieste drifters}} & 15 & 7 & 14 & 1 & 8.2 & 5.6 \\
		& 15 & 30 & 10 & 5 & 7.9 & 7.6 \\
		\hline
		\multirow{3}{*}{\textbf{Trieste HF radars}} & 225 & 10 & 202 & 23 & 10.1 & 9.9 \\
		& 225 & 50 & 112 & 113 & 9.7 & 10.2 \\
		& 225 & 90 & 23 & 202 & 8.5 & 11.2 \\
	\end{tabularx}
\end{table*}

HF-radar and drifter-based velocity measurements can exhibit differences due to distinct sampling methods, affecting both vertical and horizontal accuracy. Vertically, HF-radar velocities represent exponentially-weighted averages of the upper ocean velocity profile, influenced by the vertical shear of horizontal currents and the operating frequency of the HF-radar \cite{stewart1974hf, ivonin2004validation}. The disparity between these measurement types is further pronounced horizontally: HF-radar velocities are averaged over large grid cells spanning kilometers, whereas drifter velocities reflect motions on a scale comparable to their physical size, approximately 1 meter for CODE-type drifters \cite{enrile2018evaluation}. The reconstructed flow, obtained from both drifters and HF-radar (utilized as sampling points in this instance), exhibits a notably similar pattern with slight variations. Nevertheless, the main flow, magnitudes, direction, and RMSE values confirm the efficiency of our modeling approach that compensates for the wind, especially when a sufficient number of drifters are used to cover a substantial part of the domain. This means that any assessment without the inclusion of wind velocity would be impractical.

\subsection{Vis case}

\begin{figure*}[!h]
	\centering
	\includegraphics[width=\linewidth]{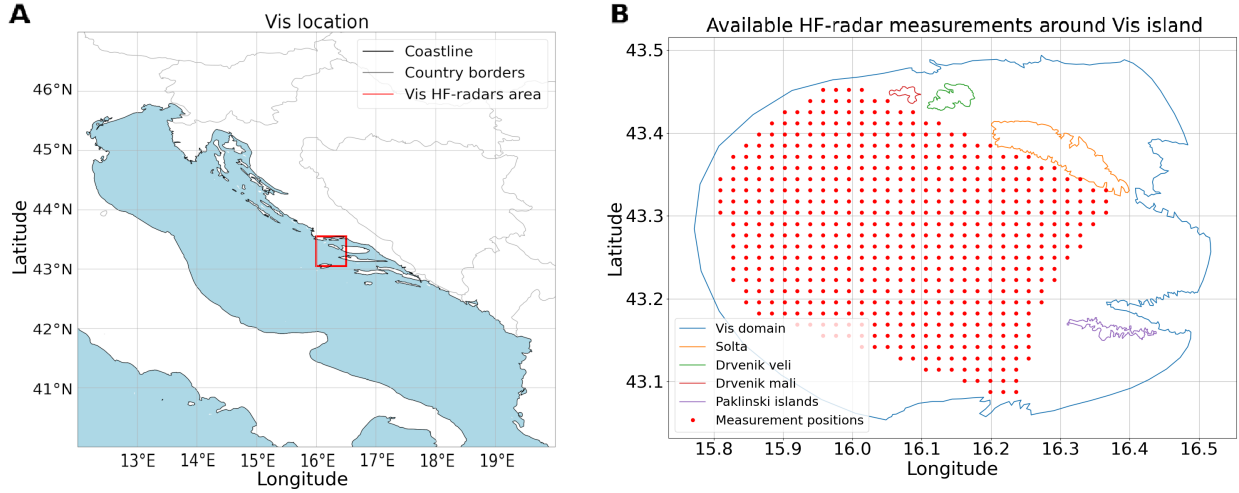}
	\caption{Figures show the location of the domain around Vis island in the Adriatic Sea and locations of available HF-radar data: (\textbf{A}) Domain around Vis island in the central part of Adriatic Sea with an area of more than 2200 square kilometers. (\textbf{B}) 555 HF-radar data values within the domain around Vis island including 4 islands and a coastline.}
	\label{fig:vis_data}
\end{figure*}

The area around the island of Vis lies in the central part of the Adriatic, close to the Croatian coast, and covers more than 2200 square kilometers. An important aspect of this case is the presence of islands within the domain, together with several inlets and outlets, making it a comprehensive test scenario covering all possible problem variants. The data for this case were obtained from the Institute of Oceanography and Fisheries \cite{izor} who provided HF data from October 2019, taken with two HF-radars that are currently inactive. We opted for a larger area than the locations of the HF-radar data as we wanted to map the evolution of the flow as it reaches the region covered by the available HF-radar data as shown in Figure \ref{fig:vis_data}.

\begin{figure*}[!h]
	\centering
	\includegraphics[width=\linewidth]{vis_results}
	\caption{Figures show surface flow reconstruction results based on HF-radar measurements at midnight on October 2nd, 2019. (\textbf{A}) HF-radar data locations with \protect\replacedRIV{corresponding velocity vectors coloured by their magnitude}{corresponding velocity magnitudes at specified time}. In this case, a larger area than the available HF-radar measurements was considered to allow the flow to develop as it reaches the measurement positions. (\textbf{B}) Illustrates the reconstructed surface flow field for a case with 277 sampling points and 278 field points (50\% reduction), where optimized velocity vectors, represented as cyan arrows, closely match the HF-measurements. This alignment signifies the successful reconstruction of the flow dynamics. (\textbf{C}) This particular case represents a very complex scenario with four islands and five different inlet/outlet boundaries. To address this complexity, 14 control points were placed on the boundaries, resulting in 28 optimization variables which is the highest number. The fitness plot displays a gradual decline from the start to the end of optimization, concluding at 1800 evaluations. Despite not reaching the desired threshold, the resemblance between HF-radar measurements and the reconstructed flow is significant.}
	\label{fig:vis_results}
\end{figure*}

\begin{table*}[!htb]
	\footnotesize
	\centering
	\caption{Data availability for the Vis case}
	\label{tab:vis_case_characteristics}
	\begin{tabularx}{\linewidth}{l@{\hspace{2cm}}rrrrrr}
		\textbf{Data Type} & Available data points & Reduction [\%] &  $n_{SP}$ &  $n_{FP}$ & $\overline{RMSE}_d$ [cm/s] & $\overline{RMSE}_f$ [cm/s] \\
		\hline
		\multirow{3}{*}{\textbf{Vis HF radars}} & 555 & 10 & 500 & 55 & 5.7 & 5.8 \\
		& 555 & 50 & 277 & 278 & 5.4 & 5.5 \\
		& 555 & 90 & 55 & 500 & 5.0 & 6.0 \\
	\end{tabularx}
\end{table*}

According to Figure \ref{fig:vis_results}, the reconstructed flow pattern \addedRIV{mostly} mirrors the flow pattern of the measurements obtained from HF-radars\addedRIV{, although some small regions with vortices are not resolved with simulation. This discrepancy can be attributed to the resolution of our simulation, which may not be adequate to capture such fine-scale features, or to the need for more measurements in this region. While we opted for faster simulations, future work could address these limitations by refining the resolution or incorporating additional data to better capture such features}. Additionally, based on 50 optimization run averages of the RMSE presented in Table \ref{tab:vis_case_characteristics}, we have demonstrated that even with a significant reduction in measurement points, the flow reconstruction remains highly accurate, validating the effectiveness of our simulation-based optimization approach in reconstructing the surface flow. Similar to the case of the Gulf of Trieste, we incorporated a wind velocity where $\mathbf{u}_w^* = [-8.25e^{-2},1.19e^{-1}]$ \protect\addedRIV{m/s} to account for atypical flows flowing directly onto the coast and around the islands. By integrating this feature, we demonstrated that our modeling approach can effectively capture real-world data and thus provide a fast approximation of the surface flow.

\section{Limitations and discussion}

The simulation-optimization methodology does not take into account factors such as wind, tides, and temperature fluctuations. This is a deliberate decision to make the model simpler and thus computationally more efficient. The dynamics of strong winds that change the direction of flow and the complicated interplay of changing tides and temperature fluctuations introduces challenges that are overlooked in the pursuit of calculation speed.

From a computational perspective, the use of coarse numerical meshes, as proposed, can be advantageous in terms of efficiency, which can be particularly useful in time-constrained search and rescue operations. However, certain disadvantages and limitations must be taken into account. The proposed simplification\addedRIV{s} may limit the ability of the surrogate to capture the complexity of a realistic flow field. \addedRIV{Inaccuracies stemming from simplifications can be addressed by employing a more detailed numerical model, however, this would certainly incur a computational penalty since calibrating advanced models to match measured results is quite challenging. Therefore, we focused on experimental data for reference, as the approach is intended to fit experimental data, with the approximated surrogate velocity field replicating real flows within an acceptable margin.} \deletedRIV{Nevertheless, it is suitable for fast approximations given that the approximated surrogate velocity field replicates real flows within an acceptable margin.} Another time-saving feature is the simulation initialization approach, which can provide significant speedup, especially in more complex domains. However, one challenge associated with the initialization approach is its potential to steer the optimization in the wrong direction, as all cases inherit the internal field from the currently best found flow field.

Problems with field reconstruction arise when there are not enough measurements points, as it is difficult to calculate field error and decide which of the four optimization outcomes our results match. This is especially true if the measurements points are confined to a small area, making it impossible to accurately reconstruct the flow in a larger domain. However, this issue can be addressed by placing drifters and using an effective number of drifters calculation to assess domain coverage and the reliability of the reconstructed flow. Fortunately, if HF-radar measurements are available for a certain area, they can be used to improve accuracy as there are usually a large number of measurements. This demonstrates the versatility of our methodology with multiple data sources.

Overall, the proposed methodology is fast and suitable for time-sensitive applications, with acceptable accuracy despite the multimodal nature of the optimization problem, assumptions and simplifications.

\section{Conclusion}

Current approaches to surface flow reconstruction, while effective in certain aspects, often lack detailed insights into the specific characteristics, behavior, and attributes defining the physical properties of a flow field. Furthermore, they can often be complex and computationally demanding, thus their applicability in real-case time-constrained scenarios is limited. In response to this limitation, a simulation-based optimization approach that utilizes a simplified two-dimensional surrogate flow model is introduced. The methodology adjusts boundary conditions in order to align the computed velocity field with scattered measurements. The intentional exclusion of some influential phenomena, such as wind, waves, changing tides, and temperature variations, prioritizes speed in our surrogate model.

From a computational perspective, the proposed use of coarse numerical meshes in our methodology improves efficiency, especially in time-sensitive search and rescue operations. Nevertheless, it is essential to acknowledge the potential inaccuracies in capturing the complexity of realistic flow fields, as our method relies on a surrogate model that omits important factors present in actual scenarios. These inaccuracies, however, are acceptable if only an approximation of the velocity field is needed. The simulation initialization approach, while providing substantial acceleration, presents challenges in the optimization process, particularly in scenarios where flow dynamics are not constrained by the coastline as is the case in the Open water scenario.

Our methodology demonstrates the effectiveness of the PSO algorithm in achieving high accuracy for reconstructing the global sea surface velocity field. The optimal number of sampling points should be tailored to each specific domain, depending on the complexity of the flow dynamics within the studied area. Here, we present a method that guides this determination, ensuring the optimal number of sampling points for any given domain. Notably, our findings emphasize the substantial improvement in reconstruction accuracy that is achieved by evenly distributing sampling points across the domain. Still, the optimal placement of drifters remains an interesting topic for further research.

Despite the simplifications, our simulation-based optimization approach brings distinct advantages over traditional methods. No extensive data, additional interpolation, or finite differencing is required, which reduces the cost and time for data collection, and provides more accurate velocity field obtainable for an entire simulated domain, even for areas with no sampling points nearby. Since drifter deployment is costly and time-consuming, our approach demonstrates that we can achieve a satisfactory approximation of the sea surface flow field with a reduced number of sampling points, unlike many studies that rely on extensive datasets. For instance, in the Gulf of Trieste case, reducing HF-radar points by 50\% \replacedRIV{led to}{resulted in} a 3\% \replacedRIV{increase}{decrease} in RMSE \addedRIV{for entire field.} \replacedRIV{In the Vis case}{, while in the Vis case}, reducing HF-radar points by 90\% achieved an RMSE of 0.060 m/s \addedRIV{for entire field}, showing a 5\% decrease in accuracy compared to the full dataset, which is consistent with the values reported in the literature. Potential field reconstruction issues due to insufficient drifters can be addressed through strategic drifter placement and effective calculations of the number of drifters. Additionally, the incorporation of HF-radar measurements, where available, enhances accuracy, showcasing the versatility of our methodology in leveraging multiple data sources.

The inclusion of a surrogate model in our simulation-based optimization approach offers a promising alternative to conventional methods, \replacedRIV{with the potential to enhance the accuracy of optimization-driven approaches within the specific context of simulation-based methodologies, especially when computational efficiency is a priority.}{and has potential to enhance accuracy and efficiency in fluid mechanics, oceanography, and environmental engineering applications.} While challenges exist, the adaptability and advantages of our methodology position it as a valuable tool for advancing research and practical applications in the field.

\section*{Acknowledgments}
The authors extend their gratitude to Riccardo Gerin, Pierre-Marie Poulain, and Milena Menna for supplying the TOSCA experiment data, and to Hrvoje Mihanović for providing HF-radar data near Vis island for the validation of this methodology. This publication is supported by the Croatian Science Foundation under the project UIP-2020-02-5090.

\section*{Data availability}
All parameters for reproducing the study are presented in the manuscript. The data needed to reproduce the presented method and cases are available on The Open Science Framework repository: \url{https://osf.io/pdtbh/}. The Python code needed to reproduce this research is available upon request.

 \bibliographystyle{elsarticle-num} 


\end{document}